\documentclass[11pt]{article} \usepackage[short]{optional}
\usepackage{fullpage,url}
\usepackage{amsmath}
\usepackage{amssymb}
\usepackage{ifpdf}
\usepackage{subfig}
\usepackage{wrapfig}
\usepackage{cite}
\usepackage{textcomp}

\opt{normal,short}{
    \usepackage{commands-tam}
    \usepackage{amsfonts}
}

\ifpdf

  \usepackage[pdftex]{epsfig}
  \usepackage[pdftex]{hyperref}

\else
    \usepackage[dvips]{epsfig}
    \newcommand{\href}[2]{#2}

\fi

\begin{document}

\title{A Domain-Specific Language for Programming in the Tile Assembly Model\footnote{This research was partially supported by NSF grants 0652569 and 0728806.}}
\author{David Doty\thanks{Department of Computer Science, Iowa State University, Ames, IA 50011, USA. ddoty@iastate.edu} \and Matthew J. Patitz\thanks{ Department of Computer
Science, Iowa State University, Ames, IA 50011, USA.
mpatitz@cs.iastate.edu.}}

\date{}

\maketitle

\begin{abstract}
We introduce a domain-specific language (DSL) for creating sets of tile types for simulations of the abstract Tile Assembly Model. The language defines objects known as tile templates, which represent related groups of tiles, and a small number of basic operations on tile templates that help to eliminate the error-prone drudgery of enumerating such tile types manually or with low-level constructs of general-purpose programming languages. The language is implemented as a class library in Python (a so-called \emph{internal DSL}), but is presented independently of Python or object-oriented programming, with emphasis on supporting the creation of visual editing tools for programmatically creating large sets of complex tile types without needing to write a program.
\end{abstract}

\section{Introduction}
\label{sec-introduction}

Erik Winfree \cite{Winf98} introduced the abstract Tile Assembly Model (aTAM) as a simplified mathematical model of molecular self-assembly. In particular, it attempts to model Seeman's efforts to coax DNA double-crossover molecules to self-assemble, programmable through the careful selection of sticky ends protruding from the sides of the double-crossover molecules. The basic component of the aTAM is the \emph{tile type}, which defines (many identical copies of) a square tile that can be translated but not rotated, which has glues on each side of strength 0, 1, or 2 (when the temperature is set to 2), each with labels, so that abutting tiles will bind with the given strength if their glue labels match. In particular, by setting the temperature to 2, we may enforce that the abutting sides of two tiles with strength-1 glues provide two input ``signals'' which determine the tile types that can attach in a given location. Such enforced cooperation is key to all sophisticated constructions in the aTAM.

\subsection{Background}
The current practice of the design of tile types for the abstract tile assembly model and related models is not unlike early machine-level programming. Numerous theoretical papers \cite{KS07,SolWin07,RotWin00,SolWin05} focus on the \emph{tile complexity} of systems, the minimum number of tile types required to assemble certain structures such as squares. A major motivation of such complexity measures is the immense level of laboratory effort required (presently) to create a physical implementation of a tile.

Early electronic computers occupied entire rooms to obtain a fraction of the memory and processing power of today's cheapest mobile telephones. This limitation did not stop algorithm developers from creating algorithms, such as divide-and-conquer sorting and the simplex method, that find their most useful niche when executed on data sets that would not have fit on all the memory existing in the world in 1950. Similarly, we hope and expect that the basic components of self-assembly will one day, through the ingenuity of physical scientists, become cheap and easy to produce, not only in quantity but in variety. The \emph{computational} scientists will then be charged with developing disciplined methods of organizing such components so that systems of high complexity can be designed without overwhelming the engineers producing the design. In this paper, we introduce a preliminary attempt at such a disciplined method of controlling the complexity of designing tile assembly systems in the aTAM.

Simulated tile assembly systems of moderate complexity cannot be produced by hand; those constructions that we have personally designed \cite{CCSA,SADSSF,SADS,RSAES} have all been produced as the output of a computer program that handles the drudgery of looping over related groups of individual tile types. However, even writing a program to produce tile types directly is error-prone, and more low-level than the ways that we tend to think about tile assembly systems.

Fowler \cite{FowlerDSLArticle} suggests that a \emph{domain-specific language (DSL)} is an appropriate tool to introduce into a programming project when the syntax or expressive capabilities of a general-purpose programming language are awkward or inadequate for certain portions of the project. Fowler distinguishes between an \emph{external DSL}, which is a completely new language designed especially for some task (such as SQL for querying and updating databases), and an \emph{internal DSL}, which is a way of ``hijacking'' the syntax of an existing general-purpose language for expressing concepts specific to the domain (such as Ruby on Rails for writing web applications). An external DSL may be as simple as an XML configuration file, and an internal DSL may be as simple as a class library. In either case the major objective is to express ``commands'' in the DSL that more closely model the way one thinks about the domain than the ``host'' language in which the project is written.

If the syntax and semantics of the DSL are precisely defined, this facilitates the creation of a \emph{semantic editor} (text-based or visual), in which programs in the DSL may be produced using a tool that can visually show semantically meaningful information such as compilation or even logical errors, and can help the programmer directly edit the abstract components of the DSL, instead of editing the source code directly. For example, Intellij IDEA and Eclipse are two programs that help Java programmers to do refactorings such as variable renaming or method inlining, which are at their core operations that act directly on the abstract syntax tree of the Java program rather than on the text constituting the source code of the program. We have kept such a tool in mind (although we have not yet implemented it) as a guide for how to appropriately structure the DSL for designing tile systems.

We structure this paper in such a way as to de-emphasize any dependence of the DSL on Python in particular or even on object-oriented class libraries in general. Eventually we will build a visual editor that mostly shields the tile set ``programmer'' from the Python roots of the implementation. The DSL provides a high-level way of \emph{thinking} about tile assembly programming, which, like any high-level language or other advance in software engineering, not only automates mundane tasks better left to computers, but also \emph{restricts} the programmer from certain error-prone tasks, in order to better guide the design, following the dictum of Antoine de Saint-Exupery that a design is perfected ``not when there is nothing left to add, but nothing left to take away.''

\subsection{Brief Outline of the DSL}
We now briefly outline the design of the tile assembly DSL. Section \ref{sec-description} provides more detail.

The most fundamental component of designing a tile assembly system manually is the tile type. In our DSL, the most fundamental component is an object known as a \emph{tile template}. A tile template represents a group of tile types (each tile type being an \emph{instance} of the tile template), sharing the same input sides, output sides, and function that transforms input signals into output signals. The two fundamental operations of the DSL are \emph{join} and \emph{add transition}. Both make the assumption that each tile template has well-defined input and output sides. In a join, an input tile template $A$ is connected to an output tile template $B$ in a certain direction $d\in\{N,S,E,W\}$, expressing that an instance $t_A$ of $A$ may have on its output side in direction $d$ an instance $t_B$ of $B$. This expresses an intention that in the growth of the assembly, $t_A$ will be placed first, then $t_B$, and they will bind with positive strength, with $t_A$ passing information to $t_B$. Whereas a join is an operation connecting the output side of a tile template to the input side of another, a transition ``connects'' input sides to output sides within a single tile template, by specifying how to compute information on the output side as a function of information on the input sides. This is called \emph{adding} a transition rather than \emph{setting}, since the information on the output sides may contain multiple independent output signals, and their computations may be specified independently of one another if convenient.

This notion of independent signals is modeled already in other DSLs for the aTAM \cite{Bec09} and for similar systems such as cellular automata \cite{ChoHuaReg02}. The join operation, however, makes sense in the aTAM but not in a system such as a cellular automaton, where each cell contains the same transition function(s). The notion of tile templates allows one to break an ``algorithm'' for assembly into separate ``stages'', each stage corresponding to a different tile template, with each stage being modularized and decoupled from other stages, except through well-defined and restricted signals passed through a join. There is a rough analogy with lines of code in a program: a single line of code may execute more than once, each time executed with the state of memory being different than the previous. Similarly, many different tile types, with different actual signal values, generated from the same tile template, may be placed during the growth of an assembly. Another difference between our language and that of \cite{Bec09} is that our language appears to be more general; rather than being geared specifically toward the creation of geometric shapes, our language is more low-level, but also more general.\footnote{An extraordinarily imprecise analogy would be that the progression \emph{manual tile programming} $\to$ \emph{Doty-Patitz} $\to$ \emph{Becker} is roughly analogous to \emph{machine code} $\to$ \emph{C} $\to$ \emph{Logo}.}


This paper is organized as follows. \opt{normal}{Section \ref{sec-prelim} provides a short formal overview of the aTAM.} Section \ref{sec-description} describes the DSL for tile assembly programming in more detail and gives examples of design and use. Section \ref{sec-conclusion} concludes the paper and discusses future work and open theoretical questions. Due to space constraints, we refer the reader to \cite{jSSADST},which contains a self-contained introduction to the Tile Assembly Model, for a formalism of the aTAM. More details and discussion may be found in \cite{Winf98,RotWin00,Roth01}. A preliminary DSL implementation can be found at \url{http://www.cs.iastate.edu/~lnsa}. 
\opt{normal}{
\section{Preliminaries}
\label{sec-prelim}
}

\newcommand{\preliminaries}{
We now give a brief description of our notation and formalism for the aTAM.  More details and discussion may be found in \cite{Winf98,RotWin00,Roth01,jSSADST}.
Our notation is that of \cite{jSSADST}, which contains a self-contained introduction to the Tile Assembly Model for the reader unfamiliar with the model.

We work in the $2$-dimensional discrete space $\Z^2$. Define the set
$U_2 = \{(0,1), (1,0), (0,-1), (-1,0)\}$ to be the set of all
\emph{unit vectors}. We
write $[X]^2$ for the set of all $2$-element subsets of a set $X$.
All \emph{graphs} here are undirected graphs, i.e., ordered pairs $G
= (V, E)$, where $V$ is the set of \emph{vertices} and $E \subseteq
[V]^2$ is the set of \emph{edges}.

Intuitively, a tile type $t$ is a unit square that can be
translated, but not rotated, having a well-defined ``side
$\vec{u}$'' for each $\vec{u} \in U_2$. Each side $\vec{u}$ of $t$
has a ``glue'' of ``color'' $\textmd{col}_t(\vec{u})$ -- a string
over some fixed alphabet $\Sigma$ -- and ``strength''
$\textmd{str}_t(\vec{u})$ -- a nonnegative integer -- specified by its type
$t$. Two tiles $t$ and $t'$ that are placed at the points $\vec{a}$
and $\vec{a}+\vec{u}$ respectively, \emph{bind} with \emph{strength}
$\textmd{str}_t\left(\vec{u}\right)$ if and only if
$\left(\textmd{col}_t\left(\vec{u}\right),\textmd{str}_t\left(\vec{u}\right)\right)
=
\left(\textmd{col}_{t'}\left(-\vec{u}\right),\textmd{str}_{t'}\left(-\vec{u}\right)\right)$.

Given a set $T$ of tile types, an \emph{assembly} is a partial
function $\pfunc{\alpha}{\Z^2}{T}$, with points $\vec{x}\in\Z^2$ at
which $\alpha(\vec{x})$ is undefined interpreted to be empty space,
so that $\dom \alpha$ is the set of points with tiles. For assemblies $\alpha$
and $\alpha'$, we say that $\alpha$ is a \emph{subassembly} of
$\alpha'$, and write $\alpha \sqsubseteq \alpha'$, if $\dom \alpha
\subseteq \dom \alpha'$ and $\alpha(\vec{x}) = \alpha'(\vec{x})$ for
all $x \in \dom \alpha$. $\alpha^\prime$ is a {\it single-tile
extension} of $\alpha$ if $\alpha \sqsubseteq \alpha^\prime$ and $\dom \alpha^\prime - \dom \alpha$ is a
singleton set.  In this case, we write $\alpha^\prime = \alpha +
(\ste{\vec{m}}{t})$, where $\{\vec{m}\} = \dom \alpha^\prime - \dom \alpha$
and $t = \alpha^\prime(\vec{m})$

A \emph{grid graph} is a graph $G =
(V,E)$ in which $V \subseteq \Z^2$ and every edge
$
\{
\vec{a},\vec{b}
\} \in E$ has the property that $\vec{a} - \vec{b} \in U_2$. The
\emph{binding graph of} an assembly $\alpha$ is the grid graph
$G_\alpha = (V, E)$, where $V =
\dom{\alpha}$, and $\{\vec{m}, \vec{n}\} \in E$ if and only if (1)
$\vec{m} - \vec{n} \in U_2$, (2)
$\color_{\alpha(\vec{m})}\left(\vec{n} - \vec{m}\right) =
\color_{\alpha(\vec{n})}\left(\vec{m} - \vec{n}\right)$, and (3)
$\strength_{\alpha(\vec{m})}\left(\vec{n} -\vec{m}\right) > 0$. 
An
assembly is $\tau$-\emph{stable}, where $\tau \in \mathbb{N}$, if every cut of $G_\alpha$ has weight at least $\tau$, where the weight of edge $\{\vec{m}, \vec{n}\}$ is $\strength_{\alpha(\vec{m})}\left(\vec{n} -\vec{m}\right)$.



A \emph{tile assembly system} (\emph{TAS}) is an ordered triple
$\mathcal{T} = (T, \sigma, \tau)$, where $T$ is a finite set of tile
types, $\sigma$ is a seed assembly with finite domain, and $\tau$ is
the temperature. In subsequent sections of this paper, we assume
that $\tau = 2$ unless explicitly stated otherwise. An
\emph{assembly sequence} in a TAS $\mathcal{T} = (T, \sigma, 2)$ is
a (possibly infinite) sequence $\vec{\alpha} = ( \alpha_i \mid 0
\leq i < k )$ of assemblies in which $\alpha_0 = \sigma$ and each
$\alpha_{i+1}$ is a single-tile extension of $\alpha_i$. The \emph{result} of an assembly sequence
$\vec{\alpha}$ is the unique assembly $\res{\vec{\alpha}}$
satisfying $\dom{\res{\vec{\alpha}}} = \bigcup_{0 \leq i <
k}{\dom{\alpha_i}}$ and, for each $0 \leq i < k$, $\alpha_i
\sqsubseteq \res{\vec{\alpha}}$. We write $\prodasm{T}$ for the
set of all \emph{producible assemblies} of $\mathcal{T}$. An
assembly $\alpha$ is \emph{terminal}, and we write $\alpha \in
\termasm{\mathcal{T}}$, if no tile can be stably added to it. We
write $\termasm{T}$ for the set of all \emph{terminal assemblies} of
$\mathcal{T}$. A TAS ${\mathcal T}$ is \emph{directed} (\emph{deterministic},
\emph{produces a unique assembly}), if $|\termasm{T}| = 1$. 
}

\opt{normal}{\preliminaries}

\section{Description of Language}
\label{sec-description}

The DSL is written as a class library in the Python programming language.  It is designed as a set of classes which encapsulate the logical components needed to design a tile assembly system in the aTAM where the temperature value $\tau = 2$.  The goal of the DSL is to abstract the tedious, low-level details of manually designing individual tile types, and thus free the developer to focus on the higher level design of larger modules.

The DSL is designed around the principle notion that data moves through an assembly as `signals' which pass through connected tiles as the information encoded in the input glues, allowing a particular tile type to bind in a location, and then, based on the `computation' performed by that tile type, as the resultant information encoded in the glues of its output edges.  (Of course, tiles in the aTAM are static objects so the computation performed by a given tile type is a simple mapping of one set of input glues to one set of output glues which is hardcoded at the time the tile type is created.)  Viewed in this way, signals can be seen to propagate as tiles attach and an assembly forms, and it is these signals which dictate the final shape of the assembly.  Using this understanding of tile-based self-assembly, we designed the DSL from the standpoint of building tile assembly systems around the transmission and processing of such signals.

We present a detailed overview of the objects and operations provided by the DSL.  We then demonstrate a full example exhibiting the way in which the DSL is used to design a tile set.

\subsection{Client-side description}

The DSL provides a collection of objects which represent the basic design units and a set of operations which can be performed on them.  We describe these objects and operations in this section, without describing the underlying data structures and algorithms implementing them.



The DSL strictly enforces the notion of input and output sides for tile types, meaning that any given side can be designated as receiving a signal (an input side), sending a signal (an output side), or neither (a blank side).

\subsubsection{DSL objects}

\paragraph{Tile system}

The highest level object is the \emph{tile system} object, which represents the full specification of a temperature $2$ tile assembly system.  It contains child objects representing the tile set and seed specification, and provides the functionality to write them out to files in a format readable by the ISU TAS simulator.

\paragraph{Tile}

In some cases, especially for tiles contained within seed assemblies, there is no computation being performed.  In such situations, it may be easier for the programmer to fully specify the glues and properties of a tile type.  The \emph{tile} object can be used to easily create such hardcoded tile types.

\paragraph{Tile template}

The principle design unit in the DSL is the tile template.  A tile template is an object which represents a collection of tile types which share the following properties:
\begin{itemize}
  \item They have exactly the same input, output, and blank sides.
  \item The types of signals received/transmitted on every corresponding input/output side are identical.
  \item The computation performed to transform the input signals to output signals can be performed by the same function, using the values specific to each instantiated tile type.
\end{itemize}

Logically, a tile template represents the set of tile types which perform the same computation on different values of the same input signal types.

\paragraph{Tile set template}

A \emph{tile set template} contains all of the information necessary for generating a tile set.  It contains the sets of tile and tile template objects which will be included in the tile set, as well as the logic for performing joins, doing error checking, etc. The tile set template object encapsulates information and operations that require communication between more than one tile template object, such as the \emph{join} operation, whereas a tile template is responsible for operations that require information entirely local to the tile template, such as \emph{add transition}.

\paragraph{Signal}

A {\it signal} is simply the name of a variable and the set of allowable values for it.  For example, a signal used to represent a binary digit could be defined by giving it the name {\it bit} and the allowable values $0$ and $1$.

\begin{figure}[htp]
\begin{center}
  \subfloat[Tansition: The function {\it calc} takes a bit and carry value as input from the bottom and right sides and computes new bit and carry values which are then output on the top and left sides.]{\label{fig:transition}\includegraphics[width=2.5in]{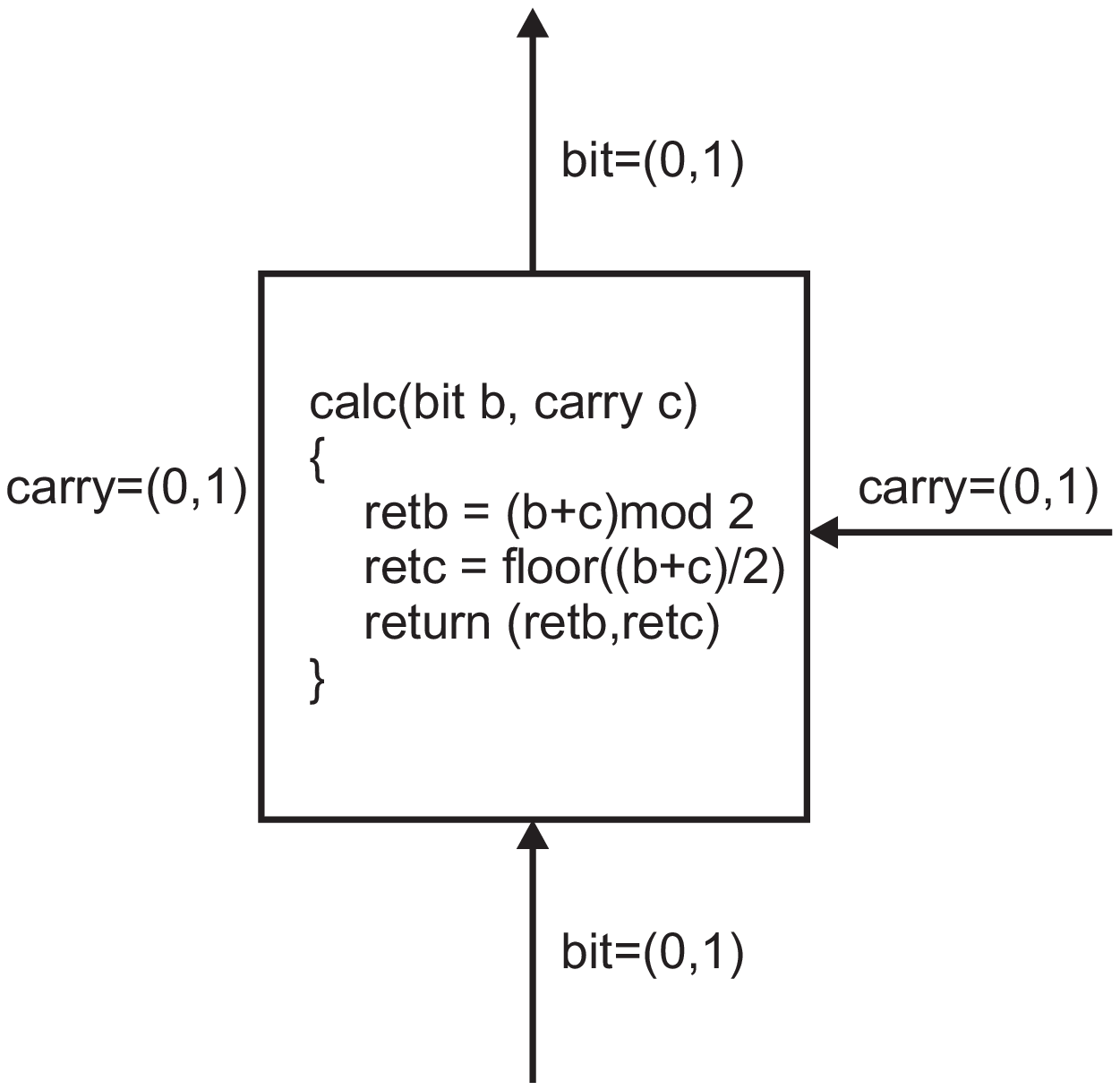}\quad}
  \quad\quad
  \subfloat[Join: The tile template on the right (tt1) is sending the signal {\it bit} with allowable values $(0,1)$ to the tile template on the left (tt2).]{\label{fig:join}\includegraphics[width=2.5in]{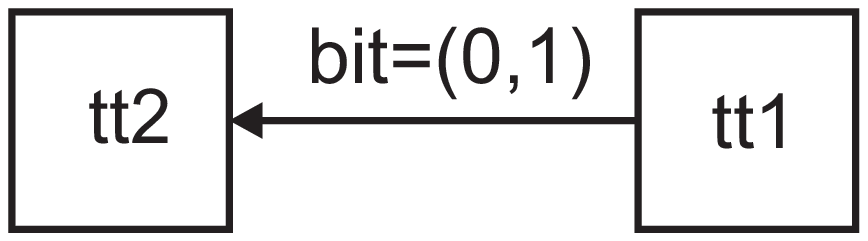}\quad\quad\quad}
  \quad\quad
  \caption{\small Logical representations of some DSL objects}
  \label{fig:DSL-objects}
\end{center}
\end{figure}

\paragraph{Transition}

{\it Transitions} are the objects which provide the computational abilities of tile templates.  A transition is defined as a set of input signal names, output signal names, and a function which operates on the input signals to yield output signals.  The logic of a function can be specified as a table enumerating all input signals and their corresponding outputs, a Python expression, or a Python function, yielding the full power of a general purpose programming language for performing the computations.  An example is shown in Figure \ref{fig:transition}.

\subsubsection{DSL operations}

\paragraph{Join}

The primary operation between tile templates, which defines the signals that are passed and the paths that they take through an assembly, is called a \emph{join}.  Joins are performed between the complementary sides (north and south, or east and west) of two tile templates (which need not be unequal), or a tile and a tile template.  If one parameter is a tile, then all signals must be given just one value; otherwise a set of possible values that could be passed between the tile templates is specified (which can still be just one).  A join is directional, specifying the direction in which a signal (or set of signals) moves.  This direction defines the input and output sides of the tile templates which a join connects.  An example is shown in Figure \ref{fig:join}.

\paragraph{Add transition}

Transition objects can be added to tile template objects, and indicate how to compute the output signals of a tile template from the input signals. If convenient, a single transition can compute more than one output signal by returning a tuple. Each output signal of a tile template with more than one possible value must have a transition computing it.

\paragraph{Add chooser}

There may be multiple tile templates that share the same input signal values on all of their input sides.  Depending on the join structure, the library may be able to use annotations (see below) to avoid ``collisions'' resulting from this, but if the tile templates share joins, then it may not be possible (or desirable) for the library to automatically calculate which tile template should be used to create a tile matching the inputs.  In this case, a user-defined function called a \emph{chooser} must be added so that the DSL can ensure that only a single tile type is generated for each combination of input values.  This helps to avoid accidental nondeterminism.

\paragraph{Set property}

Additional properties such as the display string, tile color, and tile type concentrations can be set using user-defined functions for each property.

\subsection{Additional features}

The DSL provides a number of additional, useful features to programmers, a few of which will be described in this section.  First, the DSL performs an analysis of the connection structure formed between tile templates by joins and creates \emph{annotations}, or additional information in the form of strings, which are appended to glue labels.  These annotations ensure that only tile types created from tile templates which are connected by joins can bind with each other in the directions specified by those joins, preventing the common error of accidentally allowing two tiles to bind that should not because they happen to communicate the same information in the same direction as two unrelated tiles.

Although some forms of `accidental' nondeterminism are prevented by the DSL, it does provide methods by which a programmer can intentionally create nondeterminism.  Specifically, a programmer can either design a chooser function which returns more than one output tile type for a given set of inputs, or one can add \emph{auxiliary inputs} to a tile template, which are input signals that do not come from any direction.

The DSL provides additional error-checking functionality.  For example, each tile template must have either one strength-$2$ input or two strength-$1$ inputs.  As another example, the programmer is forced to specify a chooser function if the DSL cannot automatically determine a unique output tile template, which is a common source of accidental nondeterminism in tile set design.

\subsection{Example construction}

In this section, we present an example of how to use the DSL to produce a tile assembly system which assembles a log-width binary counter.  In order to slightly simplify this example, the seed row of the assembly, representing the value $1$, will be two tiles wide instead of one.  All other rows will be of the correct width, i.e. a row representing the value $n$ will be $\ceil{\log_2(n+1)}$ tiles wide.

\begin{figure}[htp]
  \begin{center}
  \opt{normal,short}{\includegraphics[width=5.0in]{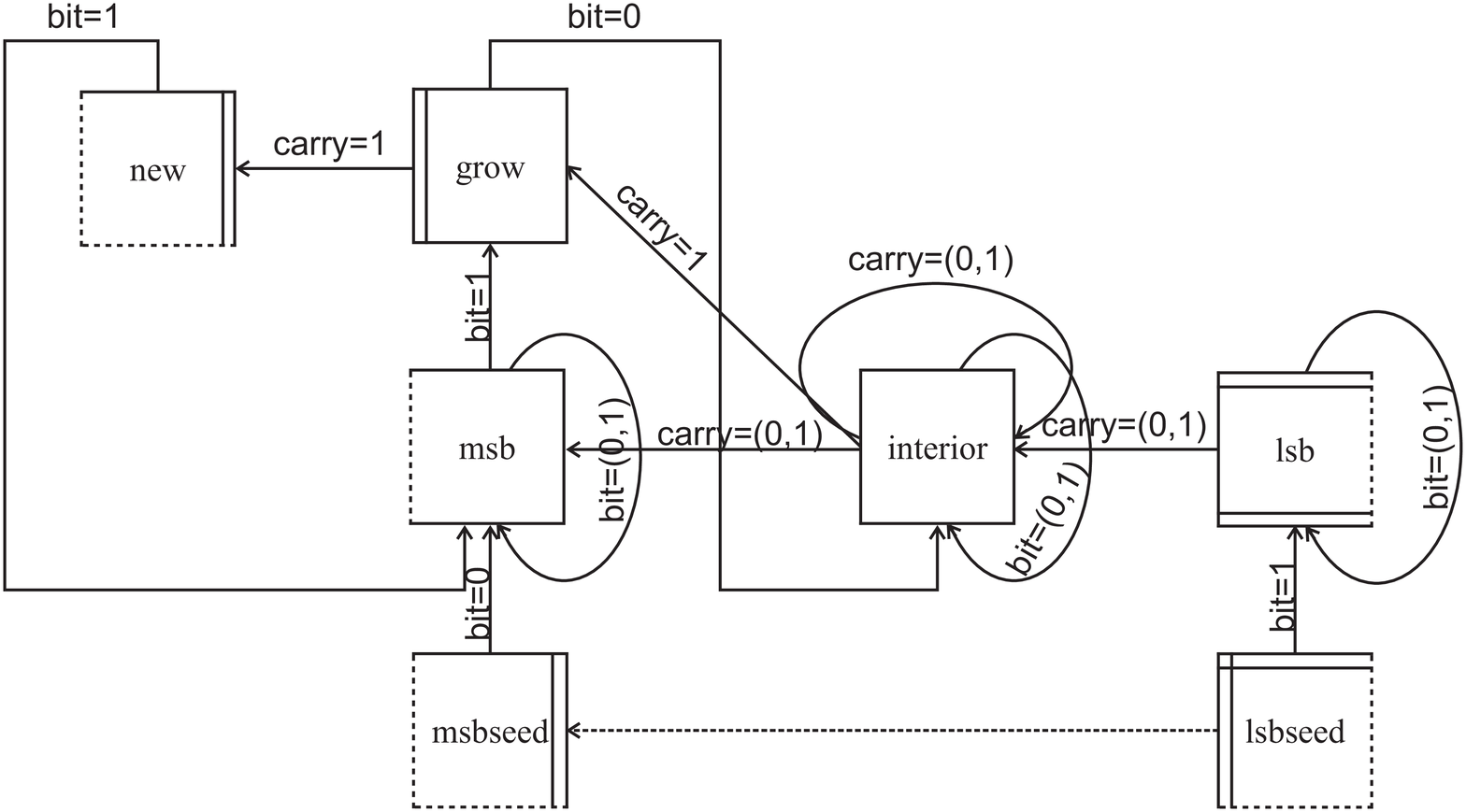}}
  \caption{\label{fig:log-width-diagram} \small Schematic diagram depicting the tile templates and joins which are used to generate a tile set that self-assembles a log-width binary counter.  All east-west joins pass the {\it carry} signal and south-north joins pass the {\it bit} signal (although some of them are restricted to subsets of the allowable values $(0,1)$).}
  \end{center}
\end{figure}

Figure \ref{fig:log-width-diagram} shows a schematic diagram representing a set of DSL objects, namely tiles, tile templates and joins, which can generate the necessary tile types.  The squares labeled {\it lsbseed} and {\it msbseed} represent hard-coded tiles for the seed row.  The other squares represent tile templates, with the names shown in the middle.  The connecting lines represent joins, which each have a direction specified by a terminal arrow and a signal which is named (either {\it bit} or {\it carry}) and has a range of allowable values (either $0$, or $1$, or both).  The dashed line between {\it lsbseed} and {\it msbseed} represents an implicit join between these two hard-coded tile types because they were manually assigned the same glues.

In addition to the joins, two more objects must be added to the tile templates to complete the necessary specifications.  Transition functions must be added to the {\it lsb}, {\it iterior}, and {\it msb} tile templates to determine how the multiple input signal values are mapped to output signal values.  Finally, since the tile templates {\it msb} and {\it grow} can both receive the signal bit=$1$ from {\it msb} and carry=$1$ from {\it interior}, a chooser function must be added to one of them so that the DSL can ensure that only a single type is generated for that situation.

The following code gives the full implementation of this example in the DSL. The first four arguments to each join method are respectively the strength, direction, ``from'' tile template, and ``to'' tile template, and the remaining arguments use Python keyword arguments to specify signals. This examples passes only one signal name through any side, but in general, multiple independent signal names can be specified.

\scriptsize
\begin{verbatim}
    lsbseed = tam.Tile(name='lsbseed', label='1', tilecolor='red', textcolor='black', westglue=('seed', 2))
    msbseed = tam.Tile(name='msbseed', label='0', tilecolor='red', textcolor='black', eastglue=('seed', 2))

    lsb = tam.TileTemplate(name='lsb')
    interior = tam.TileTemplate(name='int')
    msb = tam.TileTemplate(name='msb')
    grow = tam.TileTemplate(name='grow')
    new = tam.TileTemplate(name='new')

    tst = tam.TileSetTemplate()

    tst.join(2, N, lsbseed, lsb, bit=1)
    tst.join(1, N, msbseed, msb, bit=0)
    tst.join(2, N, lsb, lsb, bit=(0,1))
    tst.join(1, W, lsb, interior, carry=(0,1))
    tst.join(1, N, interior, interior, bit=(0,1))
    tst.join(1, W, interior, interior, carry=(0,1))
    tst.join(1, N, grow, interior, bit=0)
    tst.join(1, W, interior, msb, carry=(0,1))
    tst.join(1, W, interior, grow, carry=1)
    tst.join(1, N, msb, msb, bit=(0,1))
    tst.join(1, N, msb, grow, bit=1)
    tst.join(2, W, grow, new, carry=1)
    tst.join(1, N, new, msb, bit=1)

    def nextBitAndCarry(bit, carry):
        return ((bit + carry) % 2 , (bit + carry) // 2)

    # three ways to specify transition function
    interior.addTransition(inputs=('bit', 'carry'), outputs=('bit', 'carry'), function=nextBitAndCarry)
    lsb.addTransition(inputs=('bit'), outputs=('bit', 'carry'), table={0:(1,1), 1:(0,1))
    msb.addTransition(inputs=('bit','carry'), outputs=('bit'), expression='(bit+carry)%2')

    tst.setChooser(grow, inputs=('bit','carry'), expression='"grow" if bit == 1 and carry == 1 else "msb"')

    lsb.setLabelFunction(outputs=['bit'], inputs=[], expression='str(bit)')
    interior.setLabelFunction(outputs=['bit'], inputs=[], expression='str(bit)')
    msb.setLabelFunction(outputs=['bit'], inputs=[], expression='str(bit)')
    grow.setLabelFunction(outputs=['bit'], inputs=[], expression='str(bit)')
    new.setLabelFunction(outputs=['bit'], inputs=[], expression='str(bit)')

    tiles = tst.createTiles()
\end{verbatim}
\normalsize

\section{Conclusion and Future Work}
\label{sec-conclusion}

We have described a domain-specific language (DSL) for creating tile sets in the abstract tile assembly model. This language is currently implemented as a Python class library but is framed as a DSL to emphasize its role as a high-level, disciplined way of thinking about the creation of tile systems.

\subsection{Semantic Visual Editor}
The DSL is implemented as a Python class library, but one thinks of the ``real'' programming as the creation of visual tile templates and the joins between them, as well as the addition of signal transitions. We intend to implement a visual editor that removes the Python programming and allows the direct creation of tile templates, and the execution of the other operations, from within the editor. It will detect and report errors, such as a tile template having only one input side of strength 1, while temporarily allowing the editing to continue. Other types of errors, that do not aid the user in being allowed to persist, such as the usage of a single side as both an input and output, are prohibited outright.

One can therefore create a tile set by directly drawing the tile templates as shown in Figure \ref{fig:log-width-diagram}, while receiving helpful tips from a semantically-aware editor about errors.

\begin{wrapfigure}{r}{0.25\textwidth}
\begin{center}
  \opt{normal,short}{\includegraphics[width=1.5in]{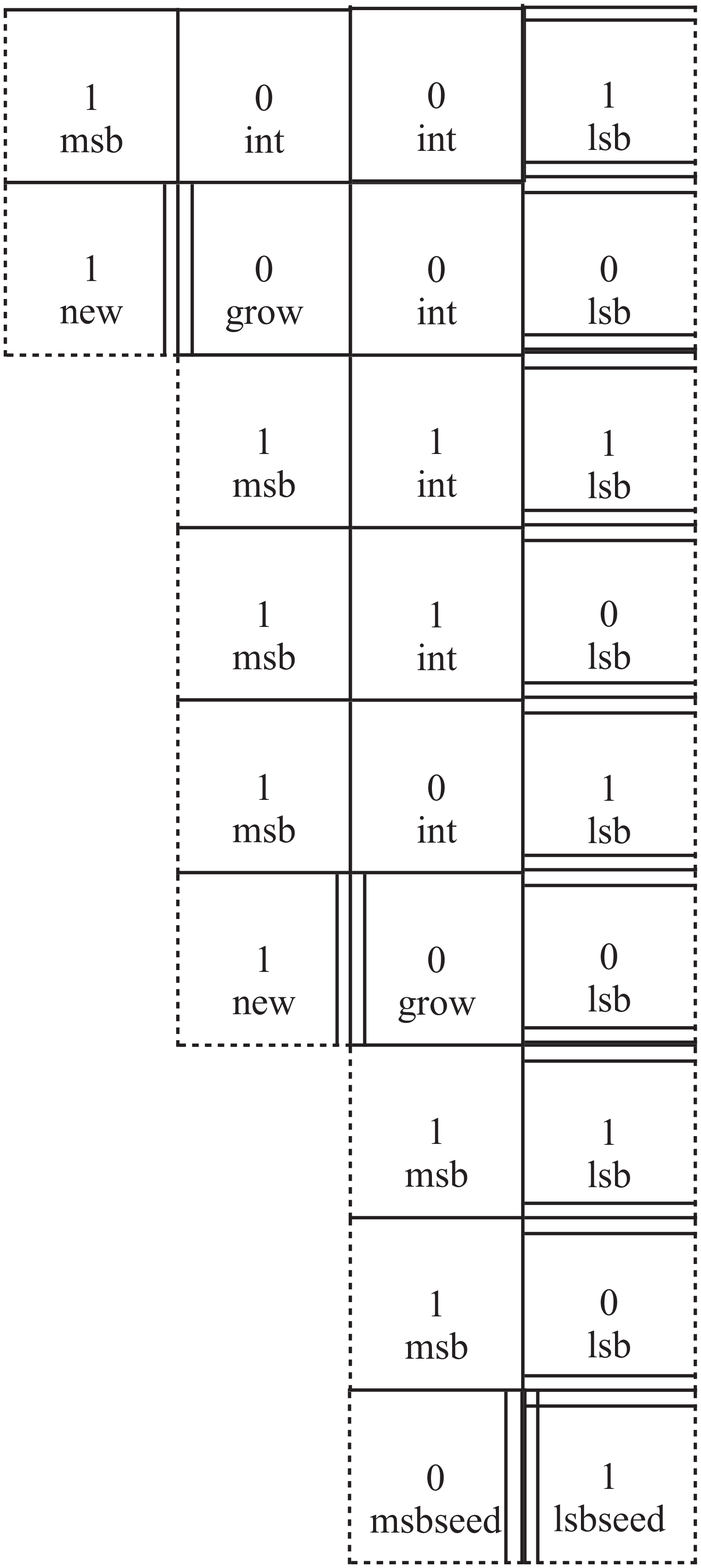}}
  \caption{\label{fig:log-width-assembly} \small The first 9 rows of the assembly of the log-width binary counter.  Note
  that each tile is labeled with its bit value on top and the name of the tile template from which it was
  generated on the bottom.}
\end{center}
\end{wrapfigure}

\subsection{Avoidance of Accidental Nondeterminism}
Initially, our hope had been to design a DSL with the property that it could be used in a straightforward way to design a wide range of existing tile assembly systems, but was sufficiently restricted that it could be guaranteed to produce a deterministic tile assembly system, so long as nondeterminism was not directly and intentionally introduced by the programmer through chooser functions or auxiliary inputs. Alas, this is not the case.

In fact, one cannot even hope for the more modest goal of stating that the DSL sufficiently restricts tile assembly systems so that the undecidable problem of detecting whether a given tile set is deterministic becomes decidable through static analysis of the join structure of the tile templates. It it straightforward to use the DSL to design a TAS that begins the parallel simulation of two copies of a Turing machine so that if the Turing machine halts, two ``lines'' of double-bonded tiles are sent growing towards each other. The tiles comprising these two lines then meet and compete nondeterministically if and only if the TM halts.

A common cause of ``accidental nondeterminism'' in the design of tile assembly systems is the use of a single side of a tile type as both an input and an output; this sort of error is prevented by the nature of the DSL in assigning each tile template unique, unchanging sets of input sides and output sides. But we cannot see an elegant way to restrict the DSL further to automatically prevent or statically detect the presence of the sort of ``geometric nondeterminism'' described in the Turing machine example. The development of such a technique would prove a boon to the designers of tile assembly systems. Conversely, perhaps there is a theorem analogous to that of Blum \cite{Blum67}, which implies that any programming language in which all programs are guaranteed to halt requires uncomputably large programs to compute some functions which are trivially computable in a general-purpose language such as Python. If this is true, then accidental nondeterminism in tile assembly, like accidental infinite loops in software engineering, would be an unavoidable fact of life, and we would do better to spend time developing formal methods for proving determinism rather than hope that it could be automatically guaranteed.

\subsection{Other Self-Assembly Models}
The abstract Tile Assembly Model is a simple but powerful tool for exploring the theoretical capabilities and limitations of molecular self-assembly. However, it is an (intentionally) overly-simplified model. Generalizations of the aTAM, such as the kinetic Tile Assembly Model \cite{Winfree98simulationsof,WinBek03}, and alternative models, such as graph-based self-assembly \cite{Klavins04}, have been studied theoretically and implemented practically. We hope to leverage the lessons learned from designing the aTAM DSL to guide the design of more advanced DSLs for high-level programming in these alternative models.

\paragraph{Acknowledgement.}
We thank Scott Summers for help testing and using preliminary versions of the DSL.

{\small
\bibliographystyle{amsplain}
\bibliography{main,dim,random,dimrelated,rbm,tam}
\clearpage
}


\end{document}